\documentclass[preprint,showpacs,preprintnumbers,amsmath,amssymb]{revtex4}

\usepackage{graphicx}
\usepackage{dcolumn}
\usepackage{bm}

\begin{document}


\title{Statistical Characterisation \& Classification of Edge
  Localised Plasma Instabilities}

\author{A. J. W\fontshape{sc}\selectfont{ebster}$^1$}
\author{R. O. D\fontshape{sc}\selectfont{endy}$^{1,2}$}
\author{JET EFDA Contributors\footnote{See the Appendix of
    F. Romanelli et al., Proceedings of the 24th IAEA Fusion Energy
    Conference 2012, San Diego, US.}}
\affiliation{JET-EFDA, Culham Science Centre, Abingdon, OX14 3DB, UK.}
\affiliation{$^1$EURATOM/CCFE Fusion Association,  
 Culham Science Centre, Abingdon, Oxfordshire, OX14 3DB, UK.}  
\affiliation{$^2$Centre for Fusion, Space and Astrophysics, Department
  of Physics, Warwick University, Coventry CV4 7AL, UK.}

\date{\today}


\maketitle

\newpage

\begin{center}
{\bf Abstract}
\end{center}

The statistics of edge-localised plasma instabilities (ELMs) in
toroidal magnetically confined fusion plasmas are considered. 
From first principles, 
standard experimentally motivated assumptions are shown to 
determine a 
specific probability distribution for the waiting times between ELMs:    
the Weibull distribution.  
This is confirmed empirically by a 
statistically rigorous  comparison with a large data set from  
the Joint European Torus (JET). 
The successful characterisation of ELM waiting times 
enables future work to progress in various ways. 
Here we present a quantitative 
classification of ELM types, 
complementary to phenomenological approaches.  
It also informs us about the nature of  ELMing processes, 
such as  
whether they are random or  deterministic.





\newpage

{\bf Introduction:} 
Edge localised plasma instabilities
(ELMs) \cite{keilhacker84,zohm96,loarte03,kamiya07}
are almost ubiquitous in high 
performance magnetically confined fusion (MCF) plasmas.   
Their phenomenological properties are correlated with the quality
of global energy confinement, and the peak energy
fluxes onto material surfaces \cite{loarte03,kamiya07,McDonald,Rapp}. 
Key challenges are to statistically characterise these processes
sufficiently well that a quantitative distinction between
different observed classes of ELMs becomes possible, and to relate
this classification to the physical 
processes responsible for them. 
This will provide a 
test for theoretical models, and is an important step towards 
improved estimates for the distribution of ELM waiting times and
sizes, both of which must be controlled in reactor-scale MCF plasma
experiments.  

ELMs offer a rich and diverse experimental 
phenomenology 
\cite{keilhacker84,zohm96,loarte03,kamiya07,McDonald,Rapp,Liang,Degeling}. 
There is intense theoretical research  
on the instabilities that may be responsible for triggering them
\cite{Webster}, 
but few unifying principles have been identified.  
We will show that widely held experimentally
motivated assumptions about ELMing require 
particular statistical characteristics. 
Specifically, if one assumes that the likelihood of ELM occurrence
increases monotonically with time elapsed since the most recent ELM, 
then the measured distribution of waiting times between
ELMs should belong to a broad class of  probability density functions
(pdfs) of which the Weibull distribution \cite{Weibull} is a special case. 
This physical approach contrasts with 
a trial and error search for a function that best fits the data
\cite{GreenhoughPoP}.  

To test this conjecture requires the identification and selection of a 
large representative data set, the development and use of a reliable 
ELM detection algorithm, and a 
method to find and compare the best possible fits between data and any
proposed pdf. 
This will provide a rigorous basis for 
present and future studies. 
As an application of our analysis, 
we distinguish
between type I and type III ELMs in a set of plasmas from the Joint
European Torus (JET) tokamak\cite{Wesson}, on the basis of 
ELM waiting time statistics alone.  
Whereas type III ELMs are usually smaller than type I ELMs, typically
they are  more frequent and the plasma's energy confinement is
lower. The ELM type is presently determined by
the ELM frequency's response to heating\cite{zohm96,loarte03,kamiya07}. 
The physically motivated derivation for our pdf 
allows a clear physical interpretation of our statistical classification.


{\bf Theoretical Background:} 
Consider the sequence and distribution of 
time intervals (waiting times) between ELMs. 
After an ELM, at $t=0$, we discuss the statistical
properties of the time of the next ELM in terms of two
linked functions.  
We define $p(t)dt$ to be the probability that the next ELM is in
the time interval $(t, t + dt)$, given that it has not yet occurred at
time $t$.  
This differs crucially from the pdf of
time intervals between ELMs, which we denote by $P(t)$, and gives the
fraction of inter-ELM time intervals that are between $t$ and $(t +
dt)$ as 
$P(t)dt$. Clearly $p(t)dt$ is a conditional probability which, multiplied
by the probability that no ELM occurs between $t = 0$ and $t$, yields the
probability  
$P(t)dt$ of an inter-ELM time interval between $t$ and $t + dt$. This gives
the identity: 
\begin{equation}\label{p1}
P(t) = p(t)  \left\{ 
1 - \int_0^t P(y) dy 
\right\}
\end{equation}
which allows $p(t)$ to be expressed in terms of $P(t)$. 
Alternately, Eq. \ref{p1} can be used to show that, 
\begin{equation}\label{bigP1}
P(t) = - \frac{d}{dt} \exp \left\{
- \int_0^t p(y) dy 
\right\}
\end{equation}
giving $P(t)$ as a function of $p(t)$,  
with $\int_0^{\infty} P(t)
dt = 1$.  
The equivalence of Eqs. \ref{p1} and \ref{bigP1} can be confirmed by
substituting Eq. \ref{bigP1} into Eq. \ref{p1}, or by writing
Eq. \ref{p1} as, $p(t)=-(d/dt) \ln \left( 1- \int_0^t P(y) dy
\right)$, and substituting into Eq. \ref{bigP1}. 


We adopt the 
experimentally motivated ansatz that  for a short time period 
$t_m$ immediately after an ELM, $p(t)=0$, 
beyond which it starts to increase. 
The simplest 
dimensionless representation of this hypothesis is,  
\begin{equation}\label{a1}
p(t) dt = \left\{
\begin{array}{lc}
0   &  t < t_m 
\\
\beta \left( \frac{t-t_m}{t_0} \right)^{\beta-1} \frac{dt}{t_0}   
& t \geq t_m
\end{array}
\right.
\end{equation}
where $t_0$ sets the time scale. 
Using Eq. \ref{bigP1}, this gives,   
\begin{equation}\label{WP}
P(t) dt = 
\left\{ 
\begin{array}{lc}
0 & t < t_m
\\ 
\beta \left( \frac{t-t_m}{t_0} \right)^{\beta-1} 
\exp \left[ - \left( \frac{t-t_m}{t_0} \right)^{\beta} \right] 
\frac{dt}{t_0} & t \ge t_m
\end{array}
\right.
\end{equation}
This is a Weibull distribution \cite{Weibull}. 
It is specified by two dimensionless 
parameters $\beta$ and $\alpha = t_m/t_0$, 
the  time scale being set by $t_0$. 
From a  theoretical perspective, the values  $\beta=1$ and
$\beta=2$ deserve special mention. 
Beyond a possible  time delay $t_m$, for $\beta=1$, $p(t)$ is
constant, 
corresponding to a ``memoryless'' process in which events occur with
equal probability independent of time. 
The transition between $p(t)$ being a concave (decreasing derivative)
and convex (increasing derivative) function is at $\beta=2$. 
As $\beta$ increases, events appear increasingly regular. 
The preceding derivation assumes 
that events are independent and that the process causing them is
stationary.



{\bf Data sets:} 
Eq. \ref{WP} will provide a good fit to a measured sequence of
waiting times when the hypothesis  represented by Eq. \ref{a1} holds. 
Such distributions have a single maximum, and require 
a macroscopic plasma equilibrium with a quasi-stationary ELMing process. 
Pdfs with additional maxima that are unlikely to have arisen
from noise were discarded, as were data whose ELM type was uncertain.  
A search of carbon-wall JET data yielded a selection 
of 70 type I and 15 type III ELM data sets. 
The data sets each have a 
steady period of ELMy H-mode lasting between $3$ and $6$ seconds, and
plasmas with 
an energy confinement time typically between $0.25$ and
$0.4$ seconds. 
The data sets 
are listed in the supplementary material (SM) \cite{SM}. 
The need for quasi-stationary ELM statistics is met by the
pulse length and quality of the JET plasmas studied, which is 
much improved on the 4 data sets studied in \cite{GreenhoughPoP}.  


{\bf ELM detection:}
ELM detection algorithms typically examine the radiation 
associated with ELMs, using a threshold in amplitude to
signal the start of an ELM, and a similar threshold or combination of
thresholds to determine when an ELM has finished \cite{GreenhoughPoP}.   
In those respects, our detection algorithm is the same. 
The advance of the algorithm 
described here is that 
the thresholds are determined from the data in a precise and
statistically invariant way, so that we do not need to reset
thresholds for different sets of data.  
This allows statistically robust comparisons between different
data sets, and  enables the technique to be used for 
non-steady-state and real-time situations if desired.  
Our algorithm  examines the signal intensity of the 
Lyman-alpha radiation from Deuterium ($D_{\alpha}$) at 
JET's inner divertor, and proceeds in two steps.
First a scan is made of the data, obtaining for each time point the 
box-average and standard deviation of the signal intensity for a time 
interval $T$ 
immediately prior to that point. 
The average and standard deviation determine a Gaussian distribution,
that is subsequently used to distinguish ELMs automatically. 
For this study the ($D_{\alpha}$) signal threshold for ELM-detection 
was for signal intensities that would  only occur 
one time in twenty, based on the
Gaussian distribution obtained from the data preceding the measurement
in question.  
Once the signal has fallen below the average again,  the ELM is
considered to have finished. 
We use a time interval $T=0.41$s that is much longer than the time
between ELMs,  but is reasonably short compared with changes to the
plasma  equilibrium. 
For stationary pulses such as those here,  with ELM waiting times  
$t \ll T$, results are 
unchanged by increasing $T$ to the time duration of the entire
dataset. 
For cases such as these, $T$ is independent of the data. 
Because we are interested in classifying ELMs by their statistical
properties, here we chose the same threshold for both the 
type I and III data. 
The threshold of one in twenty was sufficiently
sensitive for type III data, but kept noise tolerable in 
type I data. 
A systematic exploration of these thresholds will be presented elsewhere.

The method just described provides a non-subjective method
to determine when the $D_{\alpha}$ signal intensity indicates an ELM. 
Because the study involves the 
detection and study of many thousands of 
ELMs,  ``incorrect'' detection or omission of one or more ELMs
becomes part of the experimental noise. 
The detection settings require only one value to be
set in 
advance of an analysis, and because it does not need to be changed or
optimised for any given set of data, it is easy and quick to analyse
very large data sets. 
Also because thresholds are set independently of the data,
it is possible to systematically mine noisy data by varying the noise
and time-scale parameters to search for patterns in data that would
otherwise be obscured.

{\bf Best fit \& goodness of fit:}
Both the Weibull and Gaussian distributions have free parameters that
must be chosen to fit the data. 
A simple fit is provided by using the moments of the data, {\sl e.g.}
average, standard deviation, and skewness, to fit the parameters. 
More rigorously, we can consider the likelihood function for the
probability of the data given the model being considered \cite{Sivia} 
({\sl e.g.}
the Weibull model, W), and parameters $\bar{\lambda}$, with, 
\begin{equation}\label{ML}
L\left( \bar{\lambda} \right) = 
P \left( \{ t_i \} | W , \bar{\lambda} \right) 
\end{equation}
where $P(\{t_i\} | W, \bar{\lambda})$ is the probability of observing
the set of waiting 
times $\{t_i\}$, given the assumption of a Weibull
distribution (W), with fitting parameters $\bar{\lambda}$. 
The free parameters that maximise $L(\bar{\lambda})$ are their 
maximum likelihood (ML) estimate \cite{Sivia}, for which the
likelihood of the data 
(given the distribution being considered), is a maximum. 
In practice the ML estimates are found by starting from the
moment-fitted estimates and iterating to find $\bar{\lambda}$ that
maximises $L(\bar{\lambda})$. 
Given the best fits for two distributions $P_A$ and $P_B$, we can
compare their 
goodness of fit by calculating their likelihood ratio \cite{Sivia}, 
\begin{equation}\label{MLratio}
\Lambda \left( P_A , P_B \right) 
= \frac{P\left(\{t_i\} | P_A, \bar{\lambda}_A \right)}
{P\left(\{t_i\} | P_B, \bar{\lambda}_B \right)}
\end{equation} 
Under the assumption of independent $\{t_i\}$, the likelihood function
and likelihood ratio can be expanded, with for example, 
$P\left(\{t_i\} | P_A, \bar{\lambda}_A \right)=
\Pi_{i=1}^n P\left( t_i | P_A, \bar{\lambda}_A \right)$. 
Whether $P_A$ or $P_B$ is a better fit to the data is determined by
whether $\Lambda$ is greater, or less, than $1$.

Eq. \ref{WP} has one more free parameter than a Gaussian. 
Thus although Eq. \ref{WP} might provide a best fit to the data, the
model might not be better, because the
fit used an extra  parameter. 
A Bayesian analysis would introduce an extra factor \cite{Sivia} 
in Eq. \ref{MLratio} to account for 
this. 
However its influence will reduce, as the number of ELM
time intervals increases. 
Unless the factor is of order $1/\Lambda$ it will not affect the
decision for which is  the best fit. 
For the classification of
data, the most important issue is that the pdf (not the model), is a
good fit. 
From that perspective the issue is  not relevant. 
Eq. \ref{MLratio} rigorously indicates which pdf is the best 
fit, and for the large number of ELMs in our analyses,
Eq. \ref{MLratio} is  sufficient to 
determine whether the model is
significantly better or worse than a Gaussian.

\begin{figure}
\begin{center}
\includegraphics[width=7.5 cm]{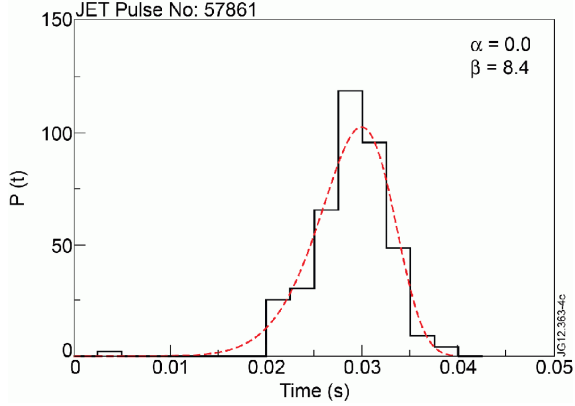}
\end{center}
\caption{ \label{fig1a}
Weibull (blue) and experimental pdfs (black bar
chart), for JET pulse no. 57861 (type I ELMs).  
}
\end{figure}

An absolute  measure of goodness of fit, is provided by 
dividing the ELM waiting time axis into intervals, calculating the
fraction $P_i$ of observed ELMs in each interval $i$, and calculating 
the co-efficient of variation 
$c_W=\langle ( P_i - P_W(t_i) )^2
\rangle/\langle P_W(t_i) \rangle^2$ between the 
 observed ($P_i$) and the 
theoretical ($P_W(t_i)$) values at the midpoint $t_i$ of the
interval. 
This gives a normalised measure of the difference between the observed
and theoretical pdfs, and provides an absolute 
measure for goodness of fit. 
It has the disadvantage of 
being dependent upon the number of data  
points used to generate the ${P_i}$. 
Small numbers of points will make $c_W$ susceptible to
noise, increasing its value. 
The choice of time intervals will also affect $c_W$, and consequently
affect a fit that minimises $c_W$.  
With enough data this would no longer be the
case, but in practice it prevents $c_W$ from determining 
a unique best fit. 
For these reasons we use a maximum likelihood best fit, which is
unique. 
Similarly if $c_W$ is used to determine which pdf gives the best
fit, the decision 
is in practice influenced by the choice of time intervals.


{\bf ELM Classification:} 
A full listing of the datasets studied, the time intervals over which
they were analysed, and the results from their analysis are presented in
the SM \cite{SM}. 
For a dataset with $n$ ELMs, we substitute Eq. \ref{WP} for $P_A$ and
a Gaussian for $P_B$ in Eq. \ref{MLratio}, then calculate the geometric
mean $\Lambda^{1/n}$ which will be of order $1$. 
If $\Lambda^{1/n}$ is greater (less) than $1.0$ then $\Lambda$ will be
much larger (smaller) for $n\gg 1$, indicating whether the Weibull is
a better (worse) fit than a Gaussian.  
For the type I datasets $\langle \Lambda^{1/n} \rangle=1.01\pm 0.04$,
where the error of  $\pm 0.04$ is  the standard deviation, 
and $n \sim 100$. 
Using time intervals of $2.5\times 10^{-3}$s, the 
coefficient of variation between the fitted and observed pdfs is  
$\langle c_W \rangle =0.63 \pm 0.22$ for the Weibull best
fits, and $\langle c_G \rangle =0.63 \pm 0.20$ for the Gaussian best fits.  
For the type III datasets  
$\langle \Lambda^{1/n} \rangle=1.51\pm 0.15$, with $n \sim 300$
or larger,   
$\langle c_W \rangle =0.70 \pm 0.23$, and  $\langle c_G \rangle=1.25
\pm 0.24$.   
Typical examples are in Figs. \ref{fig1a} and \ref{fig1b}. 
Whereas the fits are similarly good for type I ELMs, the Weibull
distribution is the clear best fit for type III ELMs.  
Substantially improved fits are likely if outliers are removed 
by improved data, improved ELM detection techniques, 
or with some algorithm.  
The values of $c_W$ and $c_G$ can be reduced if the best fit minimises 
them instead of $\Lambda$.

\begin{figure}
\begin{center}
\includegraphics[width=7.5 cm]{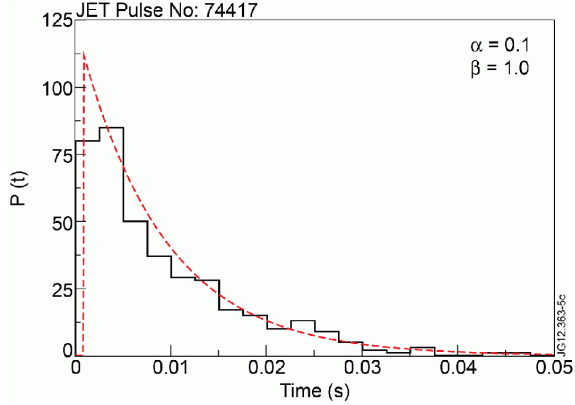}
\end{center}
\caption{ \label{fig1b}
Weibull (blue) and experimental pdfs (black bar
chart), for JET shot no. 74417 (type III ELMs).  
}
\end{figure}

Figure \ref{fig2} plots $\alpha$ and $\beta$ for the type I and type
III ELM datasets.  
There is a clear clustering of type III data for $\beta=1$ and
$\alpha < 0.5$. 
As noted earlier,  $\beta =1$ has special significance
because 
beyond an initial time delay $t_m$, it corresponds to a ``memoryless''
process in which the probability of an ELM is independent of time. 
The type I data has a wide spread in $\alpha$ and  $\beta$, but
notably $\beta$ remains of order $2$ or larger. 
As $\beta$ increases,  
ELMs will appear increasingly regular. 
Therefore the type I ELMs studied here are consistent with a process
whereby the probability of 
an ELM increases with time since the previous ELM,
possibly due to the build-up of some physical
quantity with time. 
The similarly good agreement between the Gaussian and Weibull fits
allows the alternative interpretation that type I ELMs have a specific
frequency that is broadened by noise, and that the good   
fit to type III ELM data is coincidental.
This is possible, although our original hypothesis is consistent with
present ELM models, and explains the good fit to both the
type I and  type III data.
To avoid disagreement about the classification of ELM types, our
dataset excludes ELMs whose type is uncertain. Therefore it is
possible that there is a continuum between the classifications  
that would not be observed in our data set of typical type I and type
III ELMs. 

As an example we analysed JET plasmas 66105-66109, whose ELM 
frequency is typical of  type III 
ELMs\cite{zohm96,loarte03,kamiya07,Rapp}, but whose  
$D_{\alpha}$ signal is visually similar to that of type I ELMs. 
Based on Fig. \ref{fig2},  they are not type III ELMs. 

\begin{figure}
\begin{center}
\includegraphics[width=8 cm]{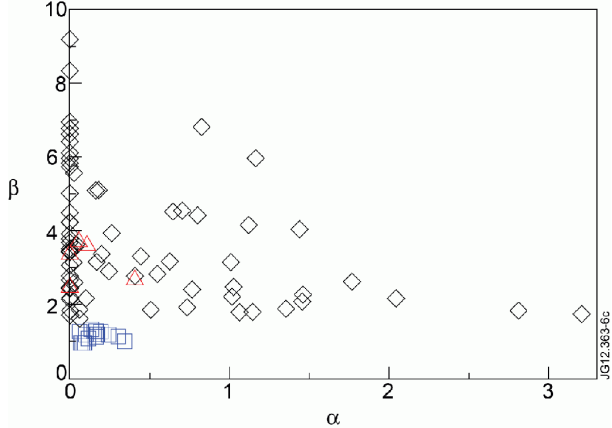}
\end{center}
\caption{ \label{fig2}
Maximum-likelihood best fits to Eq. \ref{WP}: 
type I ELM database (black diamonds), type III database (blue
squares), and some high frequency ELMs
(red triangles).
Type III data is characterised by $\beta \sim 1$, whereas all other
data has $\beta \gtrsim 2$. 
}
\end{figure}


{\bf Conclusions:} We have shown how simple     
experimentally motivated assumptions  
require 
a Weibull pdf for inter-ELM waiting times.
The model applies to stationary processes. 
A search of JET data yielded 64 sufficiently long and steady plasmas  
to test the model, 
details of which are in the SM \cite{SM}.
A statistically rigorous ELM detection technique was developed to
compare the data sets from experiments many years apart. 
The method uses a single
dimensionless threshold that is set independently of the data, and a 
single time-period, allowing  
rapid objective comparisons between
different data sets. 
The dataset was analysed, and a maximum likelihood best fit
calculated, finding a good Weibull fit    to both type I and type III data. 
Therefore we explored whether 
the dimensionless fitting co-efficients
$\alpha$ and $\beta$ could be
used to classify the data, 
concluding that they can.
The classification has 
a clear interpretation - 
type III ELMs are consistent with a memoryless 
process, but  type I ELMs  are consistent with the build-up of
a quantity with time, leading to instability.  
In contrast, present  ELM  classification 
requires either a subjective
judgment, or experimental time to determine how ELM 
frequency responds to heating 
\cite{zohm96,loarte03,kamiya07}.

To summarise, we have shown that a rigorous statistical analysis of ELM
waiting times is possible, that it can provide a quantitative 
classification of ELM types, and physical insight into the processes
responsible for them. 
The methods have numerous potential future applications, especially
for the longer plasma pulses planned for ITER\cite{ITER}. 
These include data mining, use in real-time and for other signals, and
a quantitative characterisation of the response of ELM sequences to
external parameters.

{\bf Acknowledgments:} 
Thanks to B. Alper, G. Maddison, \& M. Beurskens for
advice on ELM data.  
This work, supported by the European Communities under the contract of
Association between EURATOM and CCFE, was carried out within the
framework of the European Fusion Development Agreement. The views and
opinions expressed herein do not necessarily reflect those of the
European Commission. This work was also part-funded by the RCUK Energy
Programme under grant EP/I501045. We acknowledge the UK EPSRC for support.

\newpage 

\begin{figure}
\begin{center}
\includegraphics[width=7.5 cm]{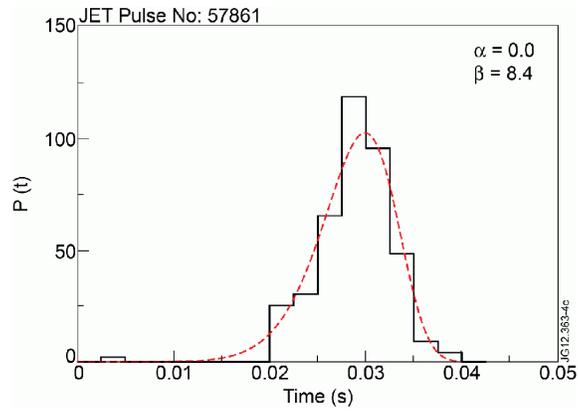}
\end{center}
\caption{ \label{fig1a}
Weibull (blue) and experimental pdfs (black bar
chart), for JET pulse no. 57861 (type I ELMs).  
}
\end{figure}


\begin{figure}
\begin{center}
\includegraphics[width=7.5 cm]{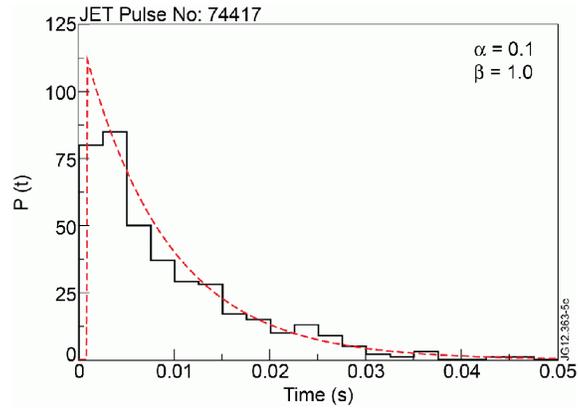}
\end{center}
\caption{ \label{fig1b}
Weibull (blue) and experimental pdfs (black bar
chart), for JET shot no. 74417 (type III ELMs).  
}
\end{figure}


\begin{figure}
\begin{center}
\includegraphics[width=8 cm]{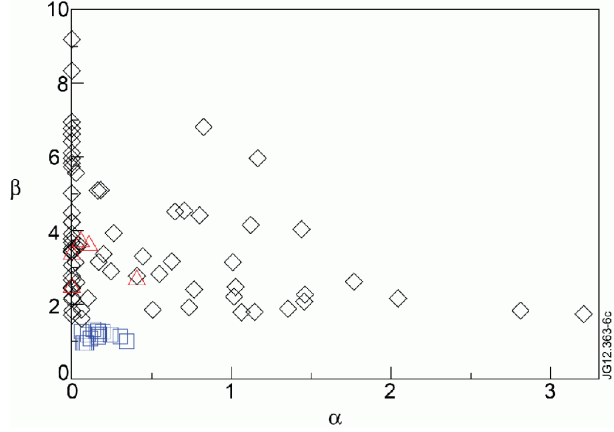}
\end{center}
\caption{ \label{fig2}
Maximum-likelihood best fits to Eq. \ref{WP}: 
type I ELM database (black diamonds), type III database (blue
squares), and some high frequency ELMs
(red triangles).
Type III data is characterised by $\beta \sim 1$, whereas all other
data has $\beta \gtrsim 2$. 
}
\end{figure}


\begin{thebibliography}{99}


\bibitem[Keilhacker et al.(1984)]{keilhacker84} M. Keilhacker, Plasma
Phys. Control. Fusion 26, 49 
(1984)


\bibitem[Zohm (1996)]{zohm96} H. Zohm, Plasma Phys. Control. Fusion
38, 105 (1996) 

\bibitem[Loarte et al.(2003)]{loarte03} A. Loarte {\em et al.}, Plasma
Phys. Control. Fusion 45, 
1549 (2003)

\bibitem[Kamiya et al.(2007)]{kamiya07} K. Kamiya {\em et al.}, Plasma
Phys. Control. Fusion 49, S43 
(2007)

\bibitem{McDonald} D.C. McDonald {\em et al.}, Fusion
  Sci. Technol. 53, 891, (2008)

\bibitem{Rapp} J. Rapp {\em et al.}, Nucl. Fusion 49, 095012, (2009)




\bibitem{Liang} Y. Liang, Fusion Sci. Technol. 59, 586, (2011)

\bibitem{Degeling} A.W. Degeling {\em et al.}, Plasma
  Phys. Control. Fusion 43, 1671, (2001)

\bibitem{Webster} A.J. Webster, Nucl. Fusion 52, 114023, (2012)

\bibitem{Weibull} W. Weibull, Transactions of the American Society of
  Mechanical Engineers, September issue, 293-297, (1951)   

\bibitem{GreenhoughPoP} J. Greenhough, S.C. Chapman, R.O. Dendy, and 
  D.J. Ward, Plasma Phys. Control. Fusion 45, 747 (2003)






\bibitem{Wesson} J. Wesson, {\sl Tokamaks} (Oxford University Press,
  Oxford, 1997) 





\bibitem{SM} See supplementary material at [] for full details of the
  datasets studied and the results of their analysis.



\bibitem{Sivia} D.S. Sivia, {\sl Data Analysis A Bayesian Tutorial}
  (Oxford University Press, Oxford, 2005)

\bibitem{ITER} R. Aymar {\em et al.} for THE ITER TEAM, Plasma
  Phys. Control. Fusion 44, 519 (2002)






















\end{thebibliography}
\end{document}


\title{Supplementary Material for Statistical Characterisation and
  Classification of Edge Localised Plasma Instabilities}

\author{A. J. W\fontshape{sc}\selectfont{ebster}$^1$}
\author{R. O. D\fontshape{sc}\selectfont{endy}$^{1,2}$}
\author{JET EFDA Contributors\footnote{See the Appendix of
    F. Romanelli et al., Proceedings of the 24th IAEA Fusion Energy
    Conference 2012, San Diego, US.}}
\affiliation{$^1$EURATOM/CCFE Fusion Association,  
 Culham Science Centre, Abingdon, Oxfordshire, OX14 3DB, UK.}  
\affiliation{$^2$Centre for Fusion, Space and Astrophysics, Department
  of Physics, Warwick University, Coventry CV4 7AL, UK.}

\pacs{52.35.Py, 05.45.Tp, 52.55Dy}

\maketitle




{\bf General Remarks:} 
An objective of this paper was to determine whether a good fit to a
large variety of data is possible by Eq. 4. 
Nonetheless, it was regarded essential that the sets have
approximately constant 
NBI heating and gas fuelling, and that they have approximately constant
central density, and energy confinement. 
As discussed in the main text, datasets whose ELM waiting time pdf
have two or more clear maxima of comparable sizes are not included. 
Some plasmas also have (approximately constant) ICRH heating during the 
time period analysed. 




In the following tables of data, the first column is the JET pulse
number, $t_1$ and $t_2$ give the time 
at which the time series analysis of $D_{\alpha}$ data started and
ended respectively, $B_T$ is the toroidal field in Tesla, $I_p$ is the
toroidal plasma current in Mega Amps, the other parameters are defined
in the main text. 



\begin{table}
\begin{center}
\normalsize
\begin{tabular}{|p{1.4cm}|p{1cm}|p{1cm}|p{2cm}|p{1cm}|p{1cm}|p{2cm}|p{1cm}
|p{1cm}|p{1cm}|p{1cm}|p{1.cm}|p{1.cm}|}
\hline
\multicolumn{13}{|c|}{\bf Type I (continued overpage)}
\\
\hline
Shot & t$_1$ & t$_2$ & $\alpha$=t$_m$/t$_0$ & $\beta$ & $c_W$ & $\mu$ &
$\sigma/\mu$ & $c_G$ & n & $\Lambda^{1/n}$ & $B_T$ & $I_p$   
\\
\hline
50564 &   62.0 &   67.0 &   0.00E+00 &   5.76 &   0.74 &   0.50E-01 &
0.19 &   0.69 & 100 &   0.98 & 1.9 & 1.9
 \\
52149 &   59.0 &   62.0 &   0.14E+01 &   4.07 &   0.70 &   0.45E-01 &
0.10 &   0.65 &  66 &   0.99 & 2.68 & 2.5
 \\
52508 &   59.5 &   63.0 &   0.21E-02 &   1.87 &   0.93 &   0.60E-01 &
0.55 &   0.95 &  57 &   1.05 & 2.6 & 2.4
 \\
52511 &   59.8 &   62.8 &   0.00E+00 &   6.97 &   0.29 &   0.31E-01 &
0.18 &   0.42 &  97 &   1.05 & 2.6 & 2.4
 \\
52513 &   59.5 &   62.8 &   0.63E+00 &   3.20 &   0.23 &   0.23E-01 &
0.20 &   0.24 & 142 &   1.01 & 2.6 & 2.4
 \\
52516 &   59.8 &   62.8 &   0.20E+00 &   3.40 &   0.36 &   0.24E-01 &
0.25 &   0.30 & 124 &   0.98 & 2.4 & 2.3
 \\
52517 &   59.8 &   62.8 &   0.10E+01 &   2.25 &   0.55 &   0.46E-01 &
0.22 &   0.59 &  64 &   1.03 & 2.4 & 2.3 
 \\
52518 &   59.8 &   62.8 &   0.10E+01 &   2.51 &   0.95 &   0.68E-01 &
0.20 &   0.95 &  43 &   1.02 & 2.4 & 2.3 
 \\
52519 &   60.7 &   63.7 &   0.00E+00 &   3.55 &   0.78 &   0.50E-01 &
0.32 &   0.76 &  59 &   0.95 & 2.4 & 2.3
 \\
52521 &   60.7 &   63.7 &   0.00E+00 &   4.25 &   0.71 &   0.50E-01 &
0.26 &   0.69 &  58 &   0.98 & 2.4 & 2.3 
 \\
53142 &   59.0 &   63.8 &   0.11E+01 &   1.84 &   0.35 &   0.27E-01 &
0.25 &   0.53 & 176 &   1.09 & 2.4 & 2.3 
 \\
56128 &   59.0 &   62.5 &   0.00E+00 &   6.44 &   0.70 &   0.39E-01 &
0.18 &   0.65 &  90 &   1.01 & 2.7 & 2.5
 \\
56143 &   59.0 &   62.0 &   0.15E+01 &   2.30 &   0.81 &   0.43E-01 &
0.17 &   0.80 &  68 &   1.03 & 2.7 & 2.5
 \\
56144 &   59.5 &   63.3 &   0.15E+01 &   2.12 &   0.61 &   0.35E-01 &
0.19 &   0.56 & 105 &   1.04 & 2.7 & 2.5
 \\
56739 &   62.5 &   67.0 &   0.65E+00 &   4.55 &   0.64 &   0.57E-01 &
0.14 &   0.59 &  78 &   0.99 & 1.4 & 1.4 
 \\
56740 &   63.5 &   67.0 &   0.14E+01 &   1.92 &   0.62 &   0.19E-01 &
0.22 &   0.74 & 185 &   1.09 & 1.4 & 1.4 
 \\
57861 &   59.0 &   63.3 &   0.00E+00 &   8.35 &   0.29 &   0.28E-01 &
0.15 &   0.49 & 150 &   1.06 & 2.7 & 2.5
 \\
57863 &   59.0 &   63.3 &   0.00E+00 &   3.45 &   0.84 &   0.41E-01 &
0.32 &   0.80 & 105 &   0.96 & 2.7 & 2.5
 \\
57865 &   59.0 &   63.3 &   0.00E+00 &   3.09 &   0.81 &   0.41E-01 &
0.35 &   0.77 & 104 &   0.96 & 2.7 & 2.5
 \\
57866 &   59.0 &   63.3 &   0.11E+01 &   1.83 &   0.43 &   0.35E-01 &
0.26 &   0.51 & 123 &   1.07 & 2.7 & 2.5
 \\
57870 &   59.0 &   63.3 &   0.26E+00 &   3.96 &   0.75 &   0.29E-01 &
0.23 &   0.78 & 146 &   1.02 & 2.7 & 2.5
 \\
57871 &   59.0 &   63.3 &   0.00E+00 &   5.04 &   0.82 &   0.30E-01 &
0.25 &   0.91 & 141 &   1.05 & 2.7 & 2.5
 \\
57872 &   59.0 &   63.3 &   0.32E+01 &   1.78 &   0.99 &   0.65E-01 &
0.13 &   1.01 &  65 &   1.09 & 2.7 & 2.5
 \\
57877 &   59.8 &   62.8 &   0.18E+01 &   2.65 &   0.49 &   0.56E-01 &
0.14 &   0.54 &  53 &   1.02 & 2.7 & 2.5
 \\
57885 &   59.0 &   62.8 &   0.40E-01 &   3.66 &   0.44 &   0.42E-01 &
0.28 &   0.41 &  89 &   0.98 & 2.7 & 2.5
 \\
57886 &   59.0 &   62.5 &   0.20E+01 &   2.20 &   0.60 &   0.48E-01 &
0.14 &   0.66 &  71 &   1.04 & 2.7 & 2.5
 \\
57888 &   59.0 &   62.8 &   0.29E-01 &   2.63 &   0.84 &   0.40E-01 &
0.35 &   0.68 &  95 &   0.97 & 2.7 & 2.5
 \\
57896 &   59.5 &   63.0 &   0.32E-01 &   3.58 &   0.64 &   0.33E-01 &
0.26 &   0.52 & 104 &   0.94 & 2.7 & 2.5
 \\
59354 &   60.0 &   63.5 &   0.73E+00 &   1.96 &   0.58 &   0.42E-01 &
0.29 &   0.62 &  83 &   1.08 & 2.7 & 2.5
 \\
60584 &   54.5 &   58.4 &   0.00E+00 &   3.48 &   0.67 &   0.54E-01 &
0.31 &   0.63 &  71 &   0.96 & 2.16 & 2.75 
 \\
60709 &   60.0 &   63.8 &   0.00E+00 &   6.64 &   0.37 &   0.33E-01 &
0.18 &   0.41 & 113 &   1.02 & 2.7 & 2.5
 \\
61471 &   59.0 &   63.5 &   0.10E+01 &   3.18 &   0.65 &   0.23E-01 &
0.15 &   0.50 & 197 &   0.97 & 2.7 & 2.5
 \\
\hline
\end{tabular}
\end{center}
\end{table}

\newpage

\begin{table}
\begin{center}
\normalsize
\begin{tabular}{|p{1.4cm}|p{1cm}|p{1cm}|p{2cm}|p{1cm}|p{1cm}|p{2cm}|p{1cm}
|p{1cm}|p{1cm}|p{1cm}|p{1.cm}|p{1.cm}|}
\hline
\multicolumn{13}{|c|}{\bf Type I (continued)}
\\
\hline
Shot & t$_1$ & t$_2$ & $\alpha$=t$_m$/t$_0$ & $\beta$ & $c_W$ & $\mu$ &
$\sigma/\mu$ & $c_G$ & n & $\Lambda^{1/n}$ & $B_T$  & $I_p$   
\\
\hline
61472 &   59.0 &   63.5 &   0.16E+00 &   5.12 &   0.23 &   0.23E-01 &
0.18 &   0.16 & 195 &   0.98 & 2.7 & 2.5
 \\
61478 &   56.7 &   59.7 &   0.10E+00 &   2.20 &   0.45 &   0.25E-01 &
0.43 &   0.46 & 120 &   1.03 & 2.5 & 3.0
 \\
61479 &   59.5 &   63.5 &   0.00E+00 &   6.80 &   0.25 &   0.24E-01 &
0.18 &   0.38 & 166 &   1.04 & 2.75 & 2.5 
 \\
61480 &   60.0 &   63.0 &   0.17E-02 &   1.75 &   0.79 &   0.31E-01 &
0.58 &   0.84 &  96 &   1.07 & 2.7 & 2.5 
 \\
62216 &   60.0 &   63.0 &   0.24E-01 &   3.17 &   0.61 &   0.34E-01 &
0.34 &   0.62 &  86 &   1.01 & 2.4 & 2.0 
 \\
62220 &   57.0 &   61.0 &   0.00E+00 &   3.46 &   0.86 &   0.61E-01 &
0.30 &   0.79 &  65 &   0.92 & 3.0 & 3.0
 \\
62221 &   57.0 &   61.0 &   0.00E+00 &   5.86 &   0.68 &   0.48E-01 &
0.19 &   0.63 &  84 &   0.98 & 3.0 & 3.0 
 \\
62222 &   57.5 &   60.5 &   0.60E-01 &   1.90 &   0.77 &   0.34E-01 &
0.51 &   0.77 &  86 &   1.04 & 3.0 & 3.0 
 \\
62224 &   57.5 &   61.0 &   0.00E+00 &   2.18 &   0.76 &   0.32E-01 &
0.48 &   0.75 & 110 &   1.02 & 3.0 & 3.0 
 \\
66111 &   58.0 &   63.0 &   0.00E+00 &   5.99 &   0.62 &   0.32E-01 &
0.22 &   0.78 & 154 &   1.05 & 2.7 & 2.5
 \\
66115 &   58.0 &   63.0 &   0.00E+00 &   3.94 &   0.85 &   0.30E-01 &
0.31 &   0.88 & 168 &   0.97 & 2.7 & 2.5
 \\
66116 &   59.0 &   63.0 &   0.83E+00 &   6.84 &   0.10 &   0.22E-01 &
0.09 &   0.21 & 184 &   1.03 & 2.7 & 2.5
 \\
67761 &   59.5 &   63.0 &   0.12E+01 &   5.99 &   0.18 &   0.15E-01 &
0.09 &   0.16 & 234 &   1.01 & 2.7 & 2.5
 \\
69373 &   63.5 &   66.5 &   0.77E+00 &   2.44 &   0.54 &   0.38E-01 &
0.23 &   0.55 &  79 &   1.02 & 1.7 & 2.0
 \\
69900 &   55.5 &   59.3 &   0.62E-01 &   1.65 &   0.67 &   0.44E-01 &
0.57 &   0.68 &  86 &   1.09 & 2.8 & 3.0 
 \\
70050 &   56.0 &   59.7 &   0.00E+00 &   2.47 &   0.56 &   0.32E-01 &
0.43 &   0.54 & 115 &   0.99 & 2.9 & 3.0 
 \\
72339 &   59.0 &   63.0 &   0.11E+01 &   4.18 &   0.54 &   0.38E-01 &
0.12 &   0.49 & 103 &   0.99 & 2.7 & 2.5
 \\
72343 &   58.5 &   63.3 &   0.00E+00 &   9.20 &   0.43 &   0.31E-01 &
0.13 &   0.32 & 155 &   1.02 & 2.7 & 2.5
 \\
72345 &   60.0 &   63.0 &   0.17E+00 &   3.19 &   1.32 &   0.26E-01 &
0.23 &   1.02 & 113 &   0.90 & 2.7 & 2.5
 \\
73087 &   59.5 &   63.3 &   0.27E-01 &   5.59 &   0.92 &   0.32E-01 &
0.18 &   0.78 & 117 &   0.95 & 2.7 & 2.5
 \\
73335 &   59.0 &   63.0 &   0.44E+00 &   3.34 &   0.84 &   0.28E-01 &
0.20 &   0.72 & 144 &   0.96 & 2.7 & 2.5
 \\
73341 &   59.0 &   63.0 &   0.28E+01 &   1.87 &   0.65 &   0.35E-01 &
0.13 &   0.90 & 114 &   1.09 & 2.7 & 2.5
 \\
73345 &   59.5 &   63.0 &   0.70E+00 &   4.57 &   0.43 &   0.34E-01 &
0.14 &   0.44 & 103 &   1.01 & 2.7 & 2.5
 \\
73346 &   59.0 &   63.0 &   0.80E+00 &   4.45 &   0.43 &   0.31E-01 &
0.13 &   0.36 & 130 &   0.98 & 2.7 & 2.5
 \\
75722 &   65.0 &   69.5 &   0.51E+00 &   1.89 &   0.74 &   0.21E-01 &
0.36 &   0.84 & 216 &   1.13 & 1.6 & 1.5 
 \\
75727 &   64.0 &   69.0 &   0.00E+00 &   4.52 &   0.91 &   0.58E-01 &
0.26 &   0.92 &  85 &   0.97 & 2.0 & 2.0 
 \\
75731 &   64.5 &   67.5 &   0.41E+00 &   2.81 &   0.71 &   0.47E-01 &
0.27 &   0.73 &  62 &   1.02 & 2.0 & 2.0
 \\
75732 &   64.5 &   67.5 &   0.25E+00 &   2.93 &   0.59 &   0.43E-01 &
0.29 &   0.59 &  69 &   1.01 & 2.0 & 2.0
 \\
76473 &   58.5 &   61.5 &   0.18E+00 &   5.13 &   0.61 &   0.39E-01 &
0.18 &   0.55 &  76 &   0.99 & 2.0 & 2.0
 \\
76474 &   58.0 &   61.5 &   0.00E+00 &   6.14 &   0.72 &   0.40E-01 &
0.19 &   0.75 &  87 &   1.03 & 2.0 & 2.0
 \\
76475 &   58.5 &   61.5 &   0.00E+00 &   3.71 &   0.79 &   0.39E-01 &
0.32 &   0.80 &  76 &   0.99 & 2.0 & 2.0
 \\
76476 &   58.5 &   61.5 &   0.00E+00 &   2.81 &   0.97 &   0.39E-01 &
0.40 &   0.94 &  76 &   1.00 & 2.0 & 2.0
 \\
\hline
\end{tabular}
\end{center}
\end{table}

\begin{table}
\begin{center}
\normalsize
\begin{tabular}{|p{1.4cm}|p{1cm}|p{1cm}|p{2cm}|p{1cm}|p{1cm}|p{2cm}|p{1cm}
|p{1cm}|p{1cm}|p{1cm}|p{1.cm}|p{1.cm}|}
\hline
\multicolumn{13}{|c|}{\bf Type I (continued)}
\\
\hline
Shot & t$_1$ & t$_2$ & $\alpha$=t$_m$/t$_0$ & $\beta$ & $c_W$ & $\mu$ &
$\sigma/\mu$ & $c_G$ & n & $\Lambda^{1/n}$ & $B_T$  & $I_p$ 
\\
\hline
76478 &   58.5 &   61.5 &   0.00E+00 &   3.81 &   0.51 &   0.38E-01 &
0.31 &   0.52 &  78 &   1.00 & 2.0 & 2.0
 \\
76479 &   58.0 &   62.0 &   0.00E+00 &   2.50 &   0.58 &   0.32E-01 &
0.42 &   0.55 & 124 &   1.00 & 2.0 & 2.0
 \\
76480 &   58.0 &   61.3 &   0.00E+00 &   2.70 &   0.44 &   0.29E-01 &
0.40 &   0.42 & 110 &   1.01 & 2.0 & 2.0
 \\
76481 &   58.0 &   61.5 &   0.00E+00 &   4.26 &   0.88 &   0.47E-01 &
0.26 &   0.85 &  73 &   0.99 & 2.0 & 2.0
 \\
76483 &   58.0 &   61.5 &   0.33E-02 &   2.22 &   0.55 &   0.29E-01 &
0.47 &   0.52 & 120 &   1.03 & 2.0 & 2.0
 \\
\hline
\end{tabular}
\end{center}
\end{table}


\begin{table}
\begin{center}
\normalsize
\begin{tabular}{|p{1.4cm}|p{1cm}|p{1cm}|p{2cm}|p{1cm}|p{1cm}|p{2cm}|p{1cm}
|p{1cm}|p{1cm}|p{1cm}|p{1.cm}|p{1.cm}|}
\hline
\multicolumn{13}{|c|}
{\bf Type III ELMs}\\
\hline
Shot & $t_1$ & $t_2$ & $\alpha = t_m/t_0$ & $\beta$ & $c_W$ & $\mu$ &
$\sigma/\mu$ & $c_G$ & n & $\Lambda^{1/n}$ & $B_T$  & $I_p$ 
\\
\hline
68608 &   62.0 &   67.0 &   0.30E+00 &   1.17 &   0.26 &   0.82E-02 &
0.68 &   0.91 & 609 &   1.39 & 2.4 & 2.0
 \\
68610 &   60.5 &   66.0 &   0.17E+00 &   1.23 &   0.60 &   0.12E-01 &
0.77 &   1.23 & 475 &   1.44  & 2.4 & 2.0
 \\
68612 &   60.5 &   66.0 &   0.62E-01 &   1.30 &   0.98 &   0.14E-01 &
0.83 &   1.35 & 395 &   1.43  & 2.4 & 2.0
 \\
68613 &   60.5 &   66.0 &   0.17E+00 &   1.32 &   0.70 &   0.14E-01 &
0.71 &   1.10 & 385 &   1.36  & 2.4 & 2.0
 \\
68614 &   60.5 &   66.0 &   0.17E+00 &   1.17 &   0.98 &   0.14E-01 &
0.90 &   1.62 & 395 &   1.65  & 2.4 & 2.0
 \\
68615 &   60.5 &   66.0 &   0.19E+00 &   1.30 &   0.83 &   0.13E-01 &
0.71 &   1.27 & 421 &   1.39  & 2.4 & 2.0
 \\
68618 &   60.5 &   66.0 &   0.13E+00 &   1.37 &   0.61 &   0.12E-01 &
0.70 &   1.08 & 445 &   1.32  & 2.4 & 2.0
 \\
68619 &   60.5 &   66.0 &   0.24E+00 &   1.22 &   0.78 &   0.13E-01 &
0.75 &   1.45 & 438 &   1.48  & 2.4 & 2.0
 \\
74410 &   56.0 &   60.5 &   0.34E+00 &   1.04 &   0.57 &   0.18E-01 &
0.74 &   1.33 & 250 &   1.53 & 2.0 & 2.5
 \\
74411 &   56.0 &   60.5 &   0.12E+00 &   1.11 &   0.51 &   0.15E-01 &
0.90 &   1.27 & 297 &   1.55 & 2.0 & 2.5
 \\
74412 &   56.0 &   60.5 &   0.77E-01 &   1.00 &   0.88 &   0.17E-01 &
0.97 &   1.14 & 263 &   1.58 & 2.0 & 2.5
 \\
74415 &   56.0 &   60.5 &   0.68E-01 &   1.00 &   0.85 &   0.14E-01 &
1.00 &   1.29 & 315 &   1.61 & 2.0 & 2.5
 \\
74417 &   57.0 &   60.5 &   0.90E-01 &   1.00 &   0.52 &   0.10E-01 &
1.08 &   1.34 & 334 &   1.77 & 2.0 & 2.5
 \\
74427 &   56.0 &   60.5 &   0.93E-01 &   1.00 &   1.06 &   0.18E-01 &
1.12 &   1.65 & 249 &   1.83 & 2.0 & 2.5
 \\
74428 &   57.0 &   60.5 &   0.10E+00 &   1.22 &   0.32 &   0.11E-01 &
0.75 &   0.71 & 313 &   1.30 & 2.0 & 2.5
 \\
\hline
\end{tabular}
\end{center}
\end{table}


\begin{table}
\begin{center}
\normalsize
\begin{tabular}{|p{1.4cm}|p{1cm}|p{1cm}|p{2cm}|p{1cm}|p{1cm}|p{2cm}|p{1cm}
|p{1cm}|p{1cm}|p{1cm}|p{1.cm}|p{1.cm}|}
\hline
\multicolumn{13}{|c|}
{\bf ``High frequency'' type I ELMs}\\
\hline
Shot & $t_1$ & $t_2$ & $\alpha = t_m/t_0$ & $\beta$ & $c_W$ & $\mu$ &
$\sigma/\mu$ & $c_G$ & n & $\Lambda^{1/n}$ & $B_T$  & $I_p$  
\\
\hline
66109 &   59.0 &   63.0 &   0.00E+00 &   3.39 &   0.62 &   0.92E-02 &
0.30 &   0.51 & 435 &   0.96 & 2.7 & 2.5
 \\
66108 &   59.0 &   62.5 &   0.00E+00 &   2.50 &   0.90 &   0.89E-02 &
0.40 &   0.80 & 390 &   0.99 & 2.7 & 2.5
 \\
66107 &   59.0 &   63.0 &   0.56E-01 &   3.73 &   0.48 &   0.83E-02 &
0.23 &   0.32 & 480 &   0.92 & 2.7 & 2.5
 \\
66106 &   59.0 &   63.0 &   0.41E+00 &   2.70 &   0.37 &   0.12E-01 &
0.26 &   0.43 & 328 &   1.01 & 2.7 & 2.5
 \\
66105 &   59.0 &   63.0 &   0.11E+00 &   3.63 &   0.48 &   0.83E-02 &
0.23 &   0.26 & 479 &   0.92 & 2.7 & 2.5
 \\
\hline
\end{tabular}
\end{center}
\end{table}

\newpage

\begin{figure}
\begin{center}
\includegraphics[width=14.6cm]{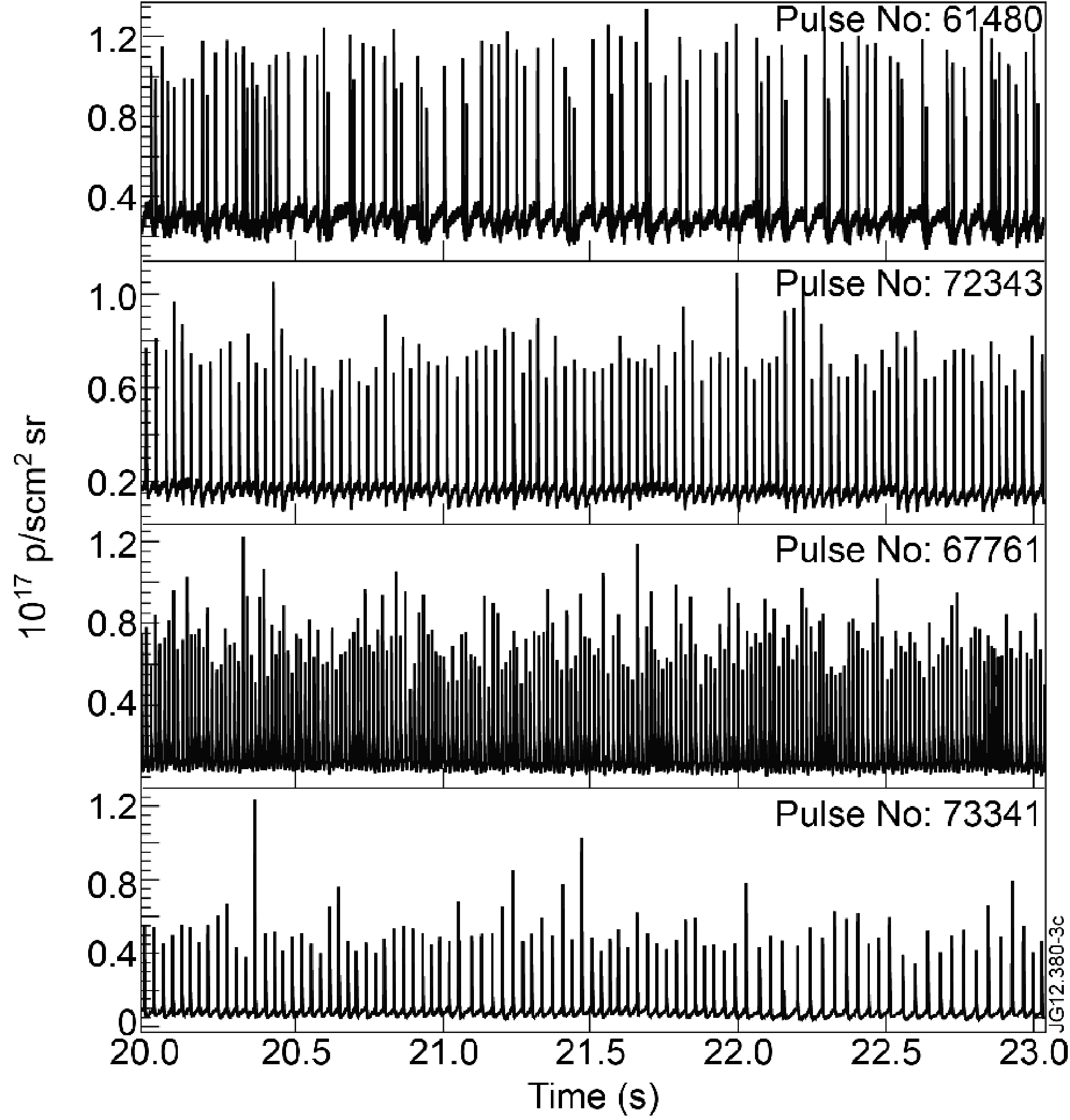}
\end{center}
\caption{ \label{Dalpha}
Example D$_\alpha$ time traces are shown for JET pulses (from the top
down): 61480 ($\alpha = 0.02$, $\beta =1.8$), 
72343 ($\alpha = 0.0$, $\beta =9.2$), 
67761 ($\alpha = 1.2$, $\beta =6.0$),
and 
73341 ($\alpha = 2.8$, $\beta =1.9$). 
}
\end{figure}
For illustrative purposes, Figure \ref{Dalpha} includes a selection of 
D$_{\alpha}$ time traces. 
The examples are chosen from the four extremities of our
$\alpha$-$\beta$ plot in Figure 3 of the main text, and are shown for
the time period of 60s-63s. The 60s-63s time window was chosen because
it is included in the analysis of all the four pulses shown.  
From top to bottom in Figure \ref{Dalpha}, or
clockwise from bottom left in the D$_\alpha$ plot of Figure 3 of the
main text, the pulses are: 
61480 ($\alpha = 0.02$, $\beta =1.8$), 
72343 ($\alpha = 0.0$, $\beta =9.2$), 
67761 ($\alpha = 1.2$, $\beta =6.0$),
and 
73341 ($\alpha = 2.8$, $\beta =1.9$).